# Girls Create Games: Lessons Learned

Bernadette Spieler, Vesna Krnjic, Wolfgang Slany
Graz University of Technology, Institute of Software Technology, Graz, Austria
bernadette.spieler@ist.tugraz.at
vesna.krnjic@ist.tugraz.at
wolfgang.slany@tugraz.at

**Abstract:** Recent studies from all over the world show that more boys than girls play video games. The numbers are different for mobile gaming apps, where 65% of women are identified as gamers. Adapting game design activities for academic purposes is a widely applied approach at schools or off-school initiatives, like CoderDojos or similar clubs, is seen as a promising opportunity for all teenagers to learn to code in an entertaining way. This raise the questions do special girls' game-design patterns exist, and what can we learn from them? This paper describes a girl-only intervention where girls were asked to create their own games. This "Girls' Coding Week" was designed as an off-school event and took place during summer 2018 with 13 girls between 11 to 14 years old. To explain the basic steps of programming and to create personalized games, the visual coding app Pocket Code, an app developed at Graz University of Technology, was used. The girls created their own games with the help of a storyboard after receiving all important information about coding (through unplugged coding activities, challenges, and a basic introduction to game design principles). Qualitative and quantitative data was collected through open interviews, as well as created artefacts and surveys which refer to motivational aspects. The findings show that gaming elements female teenagers tend to like, create, and play, mostly follow stereotypical expectations. In contrast to our experiences in heterogeneous course settings, this was not seen as something negative by girls. Furthermore, the findings provided evidence for game-making environments for girls. Subsequently, the results contributed to the development of new featured games to be used in our app to inspire female users around the world to code their own games. The authors argue that by understanding these differences in game design, we can support girls so that they become game designers and thereby more interested in coding.

## 1. Introduction

The lack of diversity in technology is a serious problem all over the world; many institutions, like the European Commission, governments, and general society recognize that this problem will affect future innovations. In engineering, manufacturing and construction-related fields, male graduates count for 72.3 percent (Baker, 2019). The "Bridging the Digital Gender Divide" report by OECD (2018) states that the gender gap is present already from an early stage and continues through university level. It is stated that girls at the age of 15 underperform boys in some ICT related skills, and only 0.5% of the girls want to pursue an ICT related career (compared to 5% of the boys). Jobs in software engineering are clearly male-dominated and women do not have a great impact on new technologies. Moreover, this means that they are not part of important decisions being made in the world of tech today and that funding will not be awarded to them to develop their ideas and concepts. Therefore, it should be in the interest of the whole computing world, rather than in the interest of any specific underrepresented group, to inspire young girls for coding (Ketelaars, 2017). In addition, there is a major lack of exposure to CS at schools all over Europe (CECE, 2017). Either schools do not offer any IT courses at all or in an inadequate amount or setting (at secondary schools it is mostly compulsory in one grade or not equally distributed over the grades), or it is an optional course.

In this paper, we first take a closer look at girl's games, gendered game design, and describe our learning tool Pocket Code. Furthermore, evidence for girls-only initiatives is summarized. Section 3 presents the research questions, the approach, and the method. In the results section, Section 4, the created artefacts (i.e., the gaming apps) are presented, as well as qualitative and quantitative data. To sum up, the discussion and conclusion sections are presented and an outlook is provided.

## 2. Literature Review

Reasons what prevent young women from choosing a career in ICT are diverse and there are general issues that must be addressed so that girls get engaged and motivated for those fields (Medel, and Pournaghshband,

2017). For example, stereotypes are present in CS and many students assume that a fanatical interest for computers and games is required in order to be successful in this field (Gabay-Egozi, Shavit, and Yaish, 2015). Inclusive environments are the key; we must consider alternative routes to IT and multiple points of entry (Frieze and Quesenberry, 2015). Playful coding initiatives designed especially for girls can support them in their decision to choose a computer science career (Zagami et al., 2015). This section presents popular games among girls, common design patterns, and subsequently, it provides arguments and characteristics of girls-only environments.

## 2.1 Girl's Games and Girl's Design

The company NewZoo (2017) published numbers of the video game industry, showing that 46% of gamers across these 13 countries are women (aged 10-65) who play on different consoles (35% PC, 48% mobile, 23% console). A percentage of 12% of women who play games are 10-20 years old. Female players hear about new games from friends/family (39%, men: 27%), social networks (20%, men 17%) or reviews/game sites and advertisements (18%, men 24-26%). Figure 1 shows the preferred genres per platform and gender.

Another statistic rated family/farm simulation and Match-3 games as the most preferred for female gamers (69% of those who are playing these genres are female) (Yee, 2017a). Match 3 games can be summarized as puzzle games where you mostly need to combine three tiles together, for example, Candy Crush Saga (Julkunen, 2015). These statistics emphasize that the genre averages range from 2% to 70%, thus developers should never focus on general statistics that consider all genders.

Ochsner (2015) collected design patterns from different girls initiatives from 1990-2007. These and literature from the same time (Agosto, 2003; Gorriz and Medina, 2000; Heeter et al., 2000) of typical design patterns, characteristics, and content girls tend to like are summarized in Figure 2.

**Figure 1:** Statistics from (NewZoo, 2017) show women prefer mostly action/adventure genres

| Common design patterns | Common game characteristics | Content similarities |
|---|---|---|
| • exploration<br>• collaboration<br>• challenge<br>• vicarious adventures<br>• sophisticated graphics and sound design<br>• role-playing<br>• realistic design | • rich narrative<br>• roles involving positive action<br>• appropriate levels of challenges<br>• opportunities to design or create<br>• engaging characters<br>• communication and collaboration<br>• use of strategies and skills | • storylines and character development<br>• real-life locales<br>• characters who are in charge of decisions and actions<br>• to create rather than to destroy<br>• involving simulation and identity play<br>• chance to swap identities |

**Figure 2:** Design, characteristics, and content among girl games.

The programming environment Scratch (https://scratch.mit.edu/) leverages game design and makes coding more accessible for a broader user group, especially novice programmers (Fields et al., 2014). A research study which observed game designs in Scratch (with a focus on racial and ethnic diversity) shows the following (Richard and Kafai., 2016): In general, most projects from female game designers focus on popular TV shows, games, or toys, or refer to mazes, dragons, and other pop culture creations and stories. Furthermore, an inspection of 52 Scratch games showed that female students used the most interactive objects that operated through mouse clicks, keystrokes, animations, stories, and projects that included multiple genres, e.g., music with interactive objects (Kafai et al., 2012).

## 2.2 The Catrobat Project: Apps to Design Personal Games

For our coding workshops, we use the learning app Pocket Code (for Android: https://catrob.at/pc, for iOS: https://catrob.at/iosPC) or Luna&Cat (https://catrob.at/luna) to explain the basic steps of programming as well as to create games. The app uses a visual programming language very similar to the one in the Scratch environment but with Pocket Code, no laptop or PC are needed; only a smartphone. In addition, Pocket Code makes access of many sensors, for instance, inclination, GPS, compass direction, etc.), and has many extensions, for example, Lego NXT/EV3 robots, drone, Arduino, or programmable embroidery machines. These apps have been developed at Graz University of Technology (TU Graz) at the Institute of Software Technology as a FOSS Open Source project with the name Catrobat (https://catrobat.org).

## 2.3 Girls-only Interventions

Existing coding club initiatives like CoderDojos (https://coderdojo.com/foundation/) have predominantly male participation (Zagami et al., 2015). To promote initiatives for female teenagers as girls-only is therefore important. These initiatives serve as vehicles to interest girls more deeply in ICT, to foster their sense of belonging and self-efficacy (Thaler and Zorn, 2010). If such activities are promoted in schools, teachers have the conflict to provide similar activities for boys as well. Moreover, situations, where females are preferred to males, can lead to a range of negative impacts (stereotypes, threats, discrimination, etc.). During the last years, researchers have come up with numerous promising approaches to get girls encouraged with coding activities. Their findings are summarized in the following (El-Nasr et al., 2007; Sadler et al., 2012; Mann and Diprete, 2013; Wang, Eccles and Kenny, 2013; Giannakos et al., 2014; Stout and Camp, 2014; Unfried et al., 2015; Zagami et al., 2015; Alvarado et al., 2017; Twentyman, 2018; Nichols, 2019):

- Improve girls self-efficacy, sense of belonging, interest, and engagement level in coding classes
- Encourage girls to create own projects which are presented in front of peers/others
- Raise girls' awareness of gender stereotypes in ICT
- Improve their expectations towards careers in programming
- Provide hands-on experiences and real-world examples
- Listen to girls suggestions about challenges and desires
- Focus on hands-on experience
- Provide early engagement (between 11 to 15) and create opportunities like extracurricular STEM activities for girls to build confidence in these areas
- Promote ICT careers and show positive role models and mentors they can both relate to and aspire to be
- Emphasize the creative aspects, idea creation, and design activities

Girls' initiatives create opportunities to focus on their interests and to enable them to socialize with other girls interested in computer science (Alvarado et al., 2017). Today, many coding clubs for girls exists, e.g., GirlsWhoCode (http://girlswhocode.com/) from the US, or Codefirst:Girls (https://www.codefirstgirls.org.uk/) from the UK.

Furthermore, facilitators and teachers report the difficulty to engage girls and boys equally in traditionally male-dominated subjects such as computing. Besides, in a school setting there are many constraints like time constraints. By default, CS in Austria is taught in two hours à 45 to 50 minutes weekly (Federal Ministry of Education, 2017). This means, that it is more difficult to convey a structuring concept over several weeks. Intensive off-school workshop weeks have the advantage that teenagers have an intensive learning period. Thus, coding initiatives for girls may improve girls's participation in such activities.

## 3. Method & Setting

The goals of the "Girls Coding Week" was twofold:
1. to provide girls with a basic set of knowledge of programming to increase their motivation, aspiration, and engagement for computer science topics (including those who do not play games), and

2. to let them design and create games in order to evaluate game design patterns used in their games to further improve girls' game design initiatives in future

For goal number 1, quantitative data was collected through questionnaires. The questionnaires (pre, daily, and post) handed out to the girls included measures of the factors regarding students' intrinsic motivators: interest, sense of belonging, self-efficacy, and engagement. The pre-questionnaires aimed to collect students' perceptions about the course, about coding, and technical fields. The daily questionnaires (handed out at the end of the first and the second double unit) had specific questions about the unit covering daily interest, fun, and achievement. Finally, a post-questionnaire asked questions about the coding week in general. For goal number 2, qualitative data was collected through interviews. Interviews were performed with all participants in groups of 2 to 4. In addition, the final gaming apps created during the course were evaluated to derive typical game design patterns from them.

The Girls Coding Week (GCW) was performed from 6th to 10th of August 2018 every day from 9 a.m. to 4 p.m. 13 girls between 11 to 14 participated in the GCW (average age: 12.8 years old). The course structure followed the PECC model (Spieler, 2018) - a gender-sensitive model that supports coding activities in the areas of Playing, Engagement, Creativity, and Coding. The workshop was conducted by two female computer science students, one female high school trainee, and the authors served as observers.

Day 1 started with an introduction game which had the goal to remember names of participants, build confidence/trust, and to get to know each other. Afterwards, participants formed groups of 4-5 members. They stayed in the same groups the whole week but they were allowed to switch groups (two did so). Each group had one facilitator allocated after each day's facilitator's switched groups. During the warm-up phase, each facilitator discussed in groups:
- Technical careers, expectations, what do programmers do, education, attributes, jobs
- What is programming, programming languages, experiences, interests
- Do you play games, apps, genres, what kind

As a result, a flipchart was created (made by the facilitator summarizing answers), which was presented and each one drew a picture of her expectation of a computer scientist. Impressions of the warm-up phase are illustrated in Figure 3.

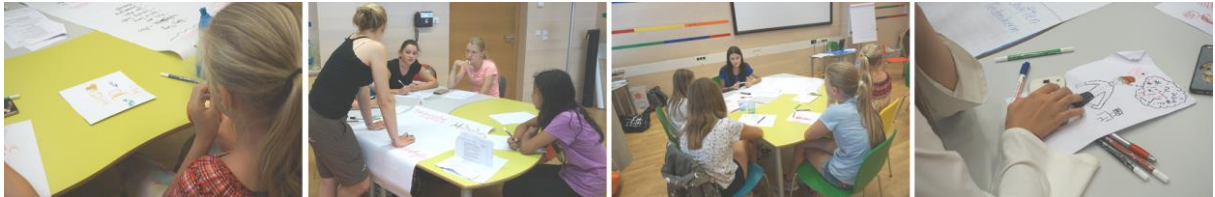

**Figure 3:** Impressions of the warm-up phase of GWC.

For the first three days, the course continued with eleven units which referred to one topic, consisting of alternating activities: a) input session, b) unplugged coding activity, and c) coding session together, d) challenge is done by everyone on their own. As a programming tool, we used our app Pocket Code. The eleven coding units are summarized in Figure 4. Between the sessions were short breaks and one bigger lunch break after three hours. At the beginning of each day, a gamified revision activity was conducted (e.g., in form of a scavenger hunt), and after the lunch and half an hour before the end of each day, we played a game together outside (e.g., Werewolf). Expressions of these phases are pictured in Figure 5.

| Unit | Topic | Unplugged Coding | Programming together | Challenge |
|---|---|---|---|---|
| Unit 0 | Objects Coordination system | Control a "robot", use coordination system | Add object, place on the screen | Introduce yourself with three objects |
| Unit 1 | Algorithm, Program, Loops | Fold a box (one does one step) | Animation | Animation with movement |
| Unit 2 | Broadcast messages | Send messages through the room | Send broadcasts when tapped | Brick challenge: use every brick |
| Unit 3 | Conditions | If you pull on the rope sth. happens | Move, glide objects with different conditions | When touched condition plus say/speak bricks |
| Unit 4 | Data types, Variables, Functions | Boxes with values that change | Create a timer | Create the game "Cookie Clicker" |
| Unit 5 | Logic, Sensors | Logic puzzles, AND or NOT | Object moves with finger position | Object moves with the inclination of phone |
| Unit 6 | Physics engine, Gravitation | Rubber ball, bouncy ball, etc. | Object react to gravity | Create a pinball game |
| Unit 7 | Pen, Stamp | One tells the other what to draw, exact coordinates | Create the patterns of a cycle, square | Create a flower or a windmill pattern |
| Unit 8 | Clones | Create a clone of a "person" | Create clones by tapping | Catch the clone game (a box catches clones) |
| Unit 9 | More bricks: Vibration, camera, flashlight | Play different games. What makes them cool? | Create a torch | Create a quiz game |
| Unit 10 | Game design | Storytelling - Red Riding Hood | Storyboard creation (graphical/textual), see next section | |

**Figure 4:** Coding units 0 -11, a) game outside, b) input session, c) unplugged coding activity, d) challenge.

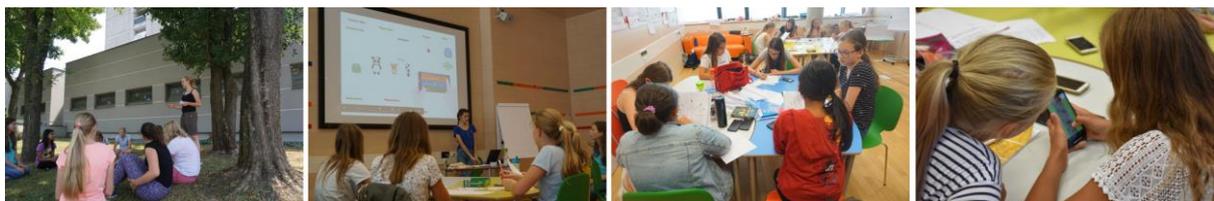

**Figure 5:** a) Games during breaks, b) presentation of the unit, c) unplugged coding, and d) programming challenges solved with the Pocket Code app

On the fourth day, the girls were introduced to two more activities. First, they had a two hours session to control Lego NXT robots with Pocket Code, and second, they were allowed to stitch the patterns made in Unit 7 with programmable embroidery machines on shirts and bags, see Figure 6.

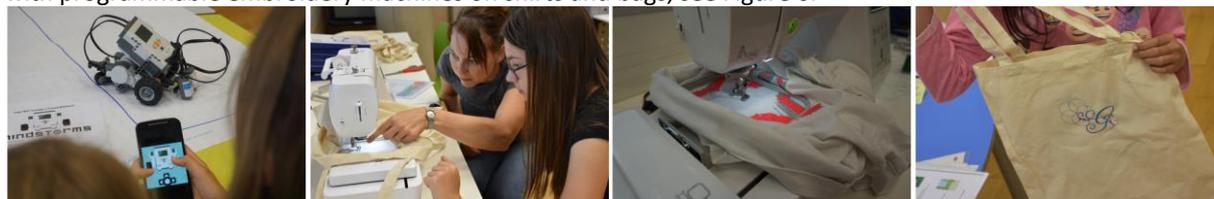

**Figure 6:** a) session with Lego NXT robots, and b) - d) stitching of patterns via an embroidery machine.

For the final gaming app, created on day four and five, the girls received the following supporting material in the form of storyboards:
- a graphical storyboard that was divided into four areas to help them to stick to the shape of a game and to frame the game in title, instruction, game, and end screen
- a textual storyboard that helped them to make important game decisions, e.g.,: name of the game, main character, gameplay (what is the game about?), genre, theme, goal of the game, used mechanics/dynamics (e.g., points, levels, difficulty levels, inventory, high-score, timer), amount of levels: what happens in level 1, 2, 3 (see Spieler and Slany, 2018a)

It was important to give participants time to think about a fitting concept for their games and about the story and what should happen. The supporting material should not only help them with these steps but also scaffold their design. All the games were collected on the last day by uploading them to our Catrobat community page. For the presentation, the girls' parents were invited. Every girl presented the game in front of the audience. They all felt very proud of their work.

## 4. Results

In reference to our first goal, during the warm-up phase (discussion in small groups) we wanted to found out more about our target group of young girls. Figure 7 summarises their discussion about their motivations to learn more about coding, their preferred apps, why they play mobile games, which programming languages they know, and what attributes the associated with women in technology.

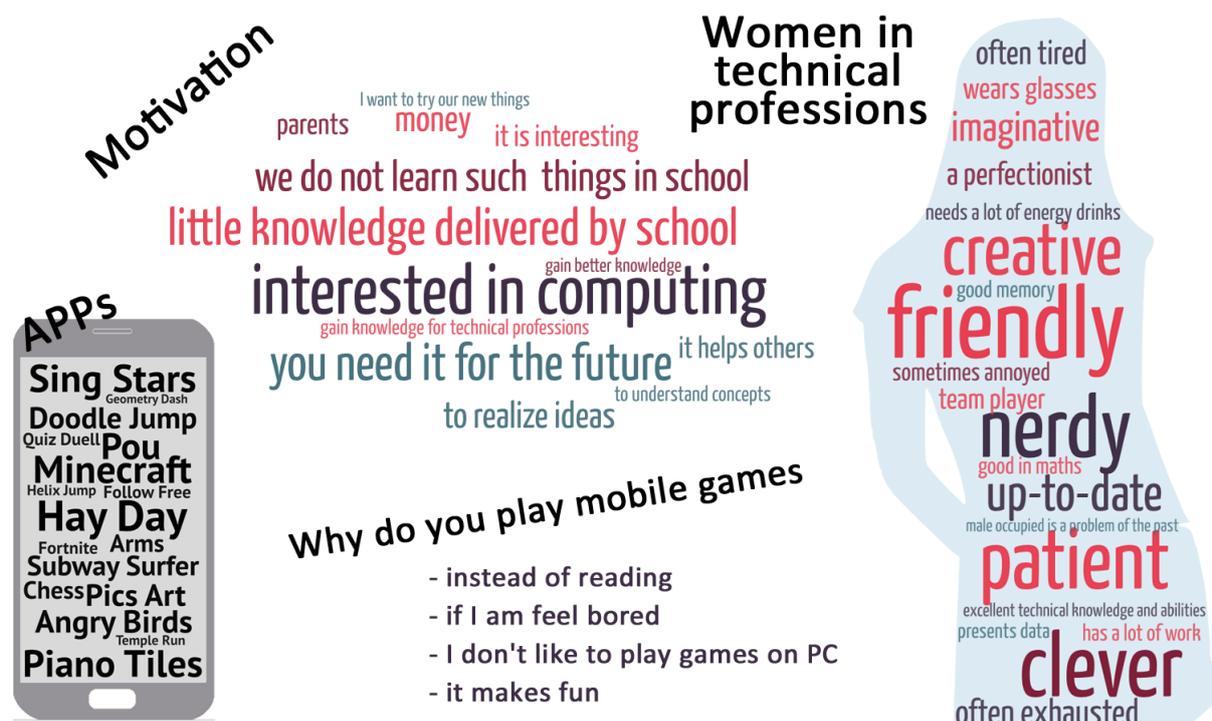

**Figure 7:** Result of the group discussions during the "Warm-Up" phase.

The questionnaires (pre, daily, and post) handed out to students included measures of the various factors regarding students' intrinsic motivators: interest, sense of belonging, self-efficacy, and fun. In all cases, a 4-point Likert scale was used to measure the variables (Sullivan and Artino, 2013). The questions have been developed at the basis of literature by Schwarzer and Jerusalem (1995), the CATS Attitude Scale Items (Krieger et al., 2015), and from other research (Li and Watson, 2011). No questions were asked that could foster stereotype threats, as proposed (Krieger et al., 2015), e.g., "Girls can do technology as well as boys". The results of all questionnaires are summarized in Figure 8. The "4 Likert Scale" refers to 1: strongly disagree, 2: disagree, 3: agree and 4: strongly agree. It is recommended to use no neutral value and to use counter questions, e.g., Coding is interesting — Coding is boring, to demand their attention (McLeod, 2008). Thus, it is not always "the higher, the better". Question with "the lower, the better" are marked with "*".

The same size of this quantitative evaluation is very small (number = 13) but should serve as a first case study for our future interventions in summer 2019.

The interest in this group of girls was stable over the whole week. There is a slight increase in self-efficacy (confidence and knowledge) over the course of the week and from day to day, they felt more proud of their achievements. Also, their sense of belonging level (e.g., knowledge) slightly increased. However, there was a slight decrease in answers pre- and post- to "coding suits me" and "coding is important". They felt slightly more engaged and had fun throughout the whole course. On average, the students agreed that they want to join similar coding courses.

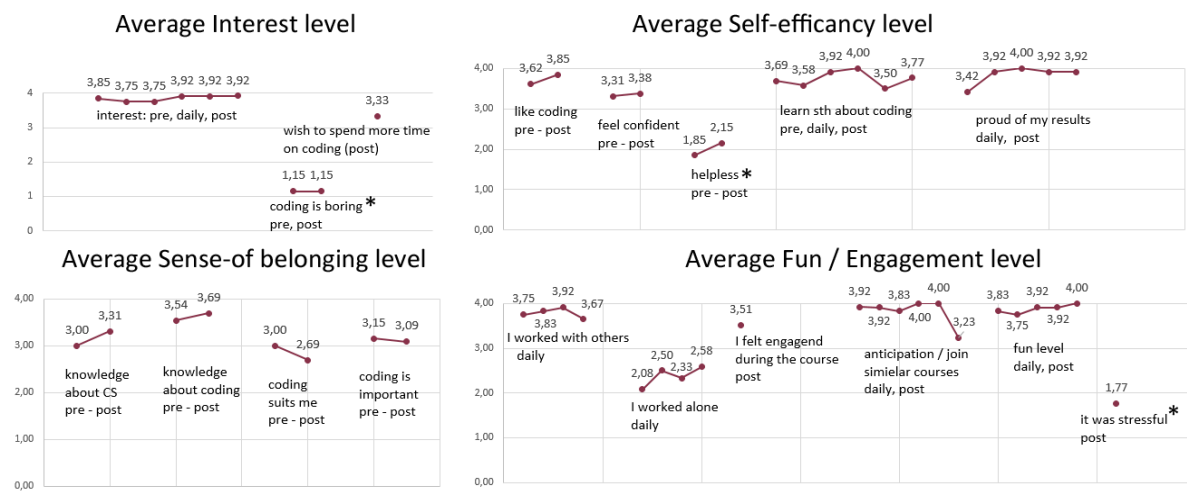

**Figure 8:** Average levels of interests, self-efficacy, sense of belonging, and fun/engagement.

The programs have been evaluated on the basis of the game design elements used (visual design, main characters) and program specific aspects (amount of scenes, objects, variables) (Spieler and Slany, 2018a). A summary of all used game elements is part of Figure 9.

| genre | theme | goals | mechanics | level of control | visual design | main character | side characters |
|---|---|---|---|---|---|---|---|
| adventure (5) simulation (4) skill game (2) action (2) | nature (6) space (2) others (realistic, horror) | catch (7) avoid (2) shoot (4) keep alive (4) | points (12) levels (12) timer (2) | inclination sensors (7) buttons (2) touch sensor (7) physics (1) combination of several sensors (6) | Catrobat media library (11) Internet (1) Paint tool (3) | animals (8) pikachu (1) monster (1) girl (2) board (1) | food (11) pokemon (1) dogs (1) boy (1) animals (2) |

**Figure 9:** Game design elements used during GCW.

The simulation games all had similar goals but used different concepts. Two of them were similar to "Tamagotchi games. The player has to feed and wash the characters, play with them, or go for a walk. One was a Pikachu simulation, the goal of which is that it evolves at the end and one was a pet simulation, where you have to care for dogs. They include "minigames", e.g., shooter, action, or catch/avoid games. The skill games were mazes and a pinball game. One of the adventure games was a text adventure which included "minigames" as well. For the main characters, girls used mostly animals like the Pocket Code family (panda, lynx, elephant, raccoon, penguin) or other animals like a dog, tiger, or horse. For side characters, all kinds of food were used, like oranges, sweets, bones, or other food for the animals to catch. Screenshots of the games are pictured in Figure 10.

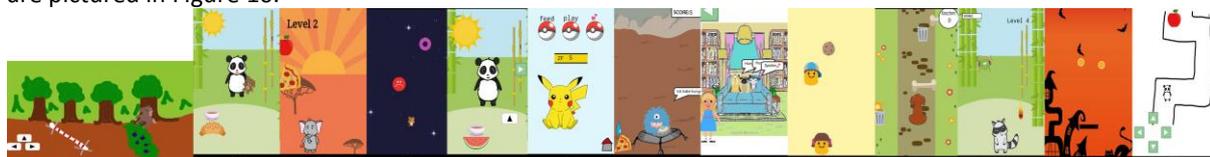

**Figure 10:** Games created during the GCW.

Regarding the programming itself, the girls used in their programs on average eight scenes (max: 17, min: 3), 74 scripts (max: 152, min: 18), 263 bricks (max: 502, min: 61), 40 objects (max: 87, min: 7), 43 looks (max: 94, min: 7), five sounds (max: 16, min: 0), and three variables (max: 14, min: 1). Summarized, all programs used various levels, included the shape of a game and numerous objects and looks, thus they were quite advanced.

During the interviews, the girls were asked out about graphics which are missing in our library. The answers included more animals from different angles (cats, horses, pets, hamster, dogs, birds, dragons, more bad animals, etc.), landscapes like forest and meadow, man-made creations such as houses and gardens, interiors (e.g., living room, kitchen, bathroom) and cities (e.g., café, shopping center, fitness center), foods such as cakes, vegetables, and fruits, or accessories like plates, smartphones, bags, shoes, or shopping clothes.

## 5. Discussion & Conclusion

The results showed that gaming elements female teenagers tend to create during the GCW mostly follow stereotypical expectations as described in Section 2.1. Compared to our previous experiences in heterogeneous groups and school classes this was not seen as something negative by girls (Spieler and Slany, 2018b).

The first aim of the paper was to provide evidence to design appropriate girls-only activities that were engaging and interesting at the same time. Results of the quantitative evaluations show the positive influence in girls' intrinsic motivation in regard to coding. This course provided a good starting point for further analysis and case studies. Even if the sample size was very small to show significant results in the quantitative evaluation, the results are still interesting. Answers to the intrinsic motivator "Self-efficiency" let us conclude that the girls had a high confidence in using the app. Furthermore, they had the feeling they learned something new and we're proud of their daily achievements. Collected factors related to the parameter "Sense of Belonging" showed that the participants learned something about technical professions and coding and hence showed a slight decrease in the feeling of the importance of coding and sense of belonging. Consequently, such a short course cannot change, for example, a strong image of stereotypes or long-held preconceptions. The fun level was high but the intention to partake in similar activities was low. Here, the study concludes that students' enjoyment has no relation to their intention to participate in similar activities again (Giannakos et al., 2014). To conclude, the quantitative evaluation shows that the values of many predictors for intrinsic motivations are located over the average. Girls agreed or strongly agreed that the coding week fostered their interests, helped them learn something about coding, helped them gain a better knowledge of technical professions and coding, and made them feel engaged while having fun during the course.

The aim of the second research question was to get more knowledge about girls design patterns. The results suggested a list of graphics be integrated into our app. New graphics (see: https://share.catrob.at/luna/media-library/looks) and featured games that have been developed on the basis of this findings together with design students from the degree program "Industrial Design" at the University of Applied Sciences in Graz (FH Joanneum). These games and some games created by the girls itself serve as feature games of the new developed Luna&Cat app. Luna&Cat is a tailored version of the app to appeal to female teenagers in particular. By showing female teenagers games designed by other young women in their age group, we help them to get ideas and inspiration to code their own programs. This is important because most girls have the feeling that the games they play are not created for them. With this customised app, our aim is to reach and build a user base of interested female teenagers who want to learn how to code.

## 6. Outlook

To engage girls in coding, a new project started in September 2018, with the name "Code'n'Stitch". With the option to program embroidery machines (very similar to the existing Turtlestitch project - https://www.turtlestitch.org/). In this way, self-created patterns and designs can be stitched on t-shirts, pants, or even bags. As a result, teenagers have something they can be proud of, something they can wear, and they can show to others. Starting in January 2019, we performed several design-thinking workshops to find out more about the requirements of the stitching extension, see Figure 11. These workshops start with a research unit, where students get asked to draw graphics which they want to stitch on clothes. Preliminary results showed that girls preferred to stitch text (sayings), flowers, hearts, and animals, and boys preferred to stitch mostly text (sayings), brands, such as Nike or Adidas.

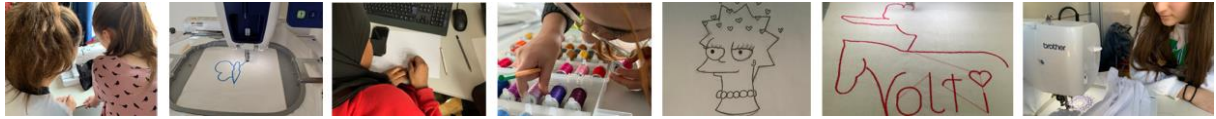

**Figure 11:** Impressions of the design-thinking workshops, Code'n'Stitch project

## Acknowledgements